# Rubidium referenced Kerr comb with cavity phase matching

Xinwei Yi†[1], Kunpeng Jia†*[1], Jingru Ji[2], Zhong Yan[3], Biaobing Jin[1,4], Zhenlin Wang[1], Wei Liang*[2], Shi-Ning Zhu[1], and Zhenda Xie*[1]

[1] *National Laboratory of Solid State Microstructures, School of Electronic Science and Engineering, College of Engineering and Applied Sciences, School of Physics, Research Institute of Superconductor Electronics (RISE) & Key Laboratory of Optoelectronic Devices and Systems with Extreme Performances of MOE, Key Laboratory of Intelligent Optical Sensing and Manipulation, Ministry of Education, and Collaborative Innovation Center of Advanced Microstructures, Nanjing University, Nanjing 210093, China*

[2] *Suzhou Institute of Nano-tech and Nano-bionics, Chinese Academy of Sciences, Suzhou 215123, China.*

[3] *Nanzhi Institute of Advanced Optoelectronic Integration, Nanjing 211800, China*

[4] *Purple Mountain Laboratories, Nanjing 211111, China*

* E-mail: jiakunpeng@nju.edu.cn; wliang2019@sinano.ac.cn; xiezhenda@nju.edu.cn

† These authors contributed equally to this work.

**Phase matching is a fundamental problem in nonlinear optics that is normally constrained by material dispersion. The limited operation wavelengths within phase matching window limits the application, including the precise metrology using Kerr combs. Demanding applications like compact optical clock and astronomical spectroscopy requires atomic reference around 800 nm, where the natural phase matching is challenging. Here we revisit the concept of cavity phase matching (CPM), and fully reveal its advantage to engineer artificial phase matching beyond material dispersion. With the access of CPM condition in a monolithic high-Q fiber Fabry-Pérot resonator featuring a macroscopic cavity length, we achieve low noise Kerr comb generation around 800 nm, within the power budget of a single-mode laser diode. Inside a pump-integrated package of 19 cm$^3$, low phase noise of -125 dBc/Hz at 100 kHz offset frequency is achieved for a 10.1 GHz repetition rate. Most importantly, the generated Kerr comb has been directly referenced to rubidium atomic transition for long-term stable operation. This result not only opens a new way for Kerr comb generation at arbitrary wavelengths, but also can be generalized to any other nonlinear optical frequency conversion application.**

Since the discovery of nonlinear optical interaction[1,2] in the 1960s, phase matching has been the crucial factor for effective nonlinear optical frequency conversion. The natural phase matching is generally limited by the material dispersion[3], and thus can only be achieved in certain nonlinear optical crystals at specific angles for specific wavelengths[4]. To address this issue, artificial phase matching[5-8] is proposed in 1962 for overcoming such material dispersion limitation. Among these artificial phase matching approaches, the cavity phase matching[9] (CPM) is elegant in principle, to operate at arbitrary wavelength for any nonlinear optical medium. By the use of a sub-coherence-length resonator, the phase correction can be achieved on the reflective mirrors among the interacting beams, and the nonlinear optical process can be enhanced while the beams recirculate. This cavity-enhanced nonlinear concept is central to modern nonlinear optics in microresonators[10,11], and naturally coexists with the phase compensation mechanism in CPM. In theory, high conversion efficiency can be achieved given the high-quality factor (Q) of the resonator, with the inherent simultaneous CPM and cavity enhancement. In practice, however, the challenge of the CPM device lies in the high-Q resonator fabrication and the phase compensation on the reflection mirrors. Previous demonstrations are limited around the natural phase matching wavelengths and requires high-peak-power pump[9].

On the other hand, Kerr frequency comb, generated in high-Q optical microresonators[10,12] and subsequently developed into dissipative soliton[13-16], has achieved significant success in recent years leveraging cavity-enhanced nonlinear frequency conversion, which offers a path toward integrated comb sources for a broad range of applications[17-23]. Within this context, Kerr combs referenced to atomic transitions in the near-infrared are of particular significance[24], as they can provide a direct and stable link to atomic frequency standards[25-28], enabling applications such as field-deployable optical clocks and precision spectroscopy. However, a significant challenge arises when entering these spectral regions as they are far from the material zero dispersion wavelength[29]; achieving phase matching via the normal geometry dispersion engineering approach presents a fundamental limitation. Intriguingly, CPM allows phase matching to be tailored independently of cavity geometry and material dispersion, thus it can in principle enable efficient frequency conversion at arbitrary wavelengths. Therefore, CPM provides a promising way to overcome the dispersion limitation and unlock practical, compact atom-referenced Kerr combs, given that the high-efficiency device can be designed and fabricated.

Here, we revisit the concept of CPM in $\chi^{(3)}$ nonlinear optical interaction and achieve Kerr comb generation at near-infrared wavelength for an example. With dielectric coating on the facets of optical fiber, CPM is realized in a sub-coherence-length fiber Fabry-Pérot resonator[30] (FFPR), enabling broad Kerr comb generation around 800 nm. With a high Q factor of $2.01\times10^{8}$ for this CPM based FFPR, Kerr comb generation is also achieved with a low threshold that is within the power budget of a single diode laser. Built in a compact self-injection-locking setup, the Kerr comb shows low phase noise of -115 dBc/Hz and -125 dBc/Hz, at 10 kHz and 100 kHz carrier frequency offsets, respectively, at its 10.1 GHz repetition rate. This result enables a direct link of the Kerr comb to the atomic transitions, such as the rubidium (Rb) $D_1$ transition lines covered in this demonstration, thereby stabilizing the comb line frequency and significantly enhancing long-term stability, a critical requirement for precision metrology applications including optical clocks. Although it is only an example of CPM in $\chi^{(3)}$ processes, this work enables Kerr comb generation at arbitrary wavelengths, thereby enabling direct referencing to abundant atomic references, a capability that paves the way for compact optical clockworks with revolutionary high-precision in time-frequency metrology.

In phase-sensitive nonlinear processes such as Kerr comb generation, efficient and broadband operation requires appropriate phase matching. This condition arises from a delicate balance among cavity dispersion, Kerr nonlinearity, and frequency detuning. To achieve phase matching over a wide bandwidth, the dispersion must remain sufficiently small, typically near the zero-dispersion point, since excessive dispersion would otherwise limit the phase-matching bandwidth. This is particularly pronounced for Kerr comb generation at visible and near-infrared wavelengths. Because these spectral regions lie beyond the natural phase matching wavelength for most $\chi^{(3)}$ mediums, which hinders the nonlinear process and suppresses Kerr comb formation. Taking the silica material for example, the zero-dispersion wavelength is around 1300 nm[31], which is quite distant from 800 nm. Here, the large material dispersion leads to severe phase mismatching, and consequently, conventional Kerr comb generation approaches[32-35] rely heavily on cavity geometry dependent dispersion engineering. This dependence on exact geometric shaping of wedge angles or coupling structure in turn introduce significant fabrication challenges such as tighter tolerances and greater susceptibility to defects. Given these practical limitations, it therefore becomes meaningful to reconsider the problem from the perspective of phase mismatch compensation, rather than relying on dispersion engineering.

Interestingly, even under such high dispersion caused phase mismatching, directional frequency conversion can still occur within a limited frequency range over a macroscopic length shorter than the coherence length $L_{coh}$. This is exploitable in a cavity only when its length $L_{cav}$ is shorter than $L_{coh}$ so that the nonlinear conversion proceeds directionally and backward conversion is avoided. In this condition, phase matching ensures efficient nonlinear conversion each round trip. Considering degenerate four wave mixing process, this coherence length $L_{coh}$ can be expressed as:

$$L_{coh} = \pi/\Delta k = \pi/(2k_p-k_s-k_i), \quad (1)$$

where $\Delta k$ is the wave vector mismatching, and $k_p$, $k_s$ and $k_i$ represent wave vectors for the pump, signal and idler, respectively. Previously, all the experimental demonstrations of CPM are based on $\chi^{(2)}$ processes. Considering the octave-spanning wavelength difference, the coherence length in $\chi^{(2)}$ processes is typically on the order of a few micrometers level, and critically, all $\chi^{(2)}$-based CPM operates near the natural phase-matching points. Consequently, CPM in this context does not effectively extend the phase-matching bandwidth, which is the most fundamental limitation. Therefore, $\chi^{(2)}$-based CPM also suffers from a high power threshold and is unable to operate with continuous-wave lasers. In contrast, the Kerr comb generation process based on $\chi^{(3)}$ nonlinearity in this work exhibits a significantly longer coherence length than that of $\chi^{(2)}$. Furthermore, thanks to the ultra-low propagation loss of fiber materials, high-Q resonators can be constructed to achieve substantial cavity enhancement, enabling low threshold and efficient frequency conversion.

In our FFPR design as shown in Fig. 1a, we keep the cavity length $L_{cav}$ smaller than $L_{coh}$ and introduce phase compensation on the cavity mirrors. This compensation rectifies the accumulated phase mismatching after every reflection, thereby realigning the phase of optical waves on the cavity mirrors. This mechanism is CPM, which functions to coherently extend directional nonlinear conversion from the sub-coherence scale to the macroscopic scale. As a result, efficient nonlinear frequency conversion occurs continuously over multiple roundtrips, with the effective interaction length of optical waves far exceeding $L_{coh}$.

This ability to maintain efficient conversion over macroscopic lengths is quantified with the relationship between $\Delta\omega$ (frequency detuning from signal/idler to pump) and $L_{coh}$ in the fiber (Fig. 1b), which is critical for FFPR fabrication. We find that coherence length is as long as 60 mm for $\Delta\omega$ of 36 THz. Such a macro dimension is beneficial for high-Q FFPR fabrication, considering the ultra-low loss of optical fiber. To simplify the mirror coating design (Supplementary Fig. S3) in this first $\chi^{(3)}$ CPM demonstration, we choose an $L_{cav}$ of 10 mm which is smaller than $L_{coh}$. We calculate the phase mismatching in single trip, $\Delta\varphi = \Delta k \times L_{cav}$, as a function of frequency detuning $\Delta\omega$ from the pump wavelength around 800 nm. The black curve (Fig. 1b) shows the phase mismatching that accumulates in the optical fiber, which originates from the material dispersion of silica. The gray curve (Fig. 1b) represents the additional phase introduced by the dielectric coating mirror, which is specifically designed to counteract the phase mismatching accumulated in the fiber. The red curve (Fig. 1b) corresponds to the overall phase mismatching of the FFPR, which is significantly suppressed across a broader spectral bandwidth due to the complementary compensation between the fiber and the coating. Based on this configuration, we fabricate the FFPR (see Methods) from few-mode fiber and coating mirrors. Each mirror induces a phase modulation that compensates the phase mismatching accumulated during a single propagation of 10 mm (Supplementary Fig. S1). Although phase mismatching accumulates with propagation after each light reflection on end mirrors, demonstrating CPM as a non-strict phase matching approach, efficient nonlinear frequency conversion still occurs since $\Delta\varphi$ remains below $\pi$.

In conventional Kerr comb generation, different states can be generated in cavities with different dispersion[14,36-38], along with the combined influence of Kerr nonlinearity and pump frequency detuning. In our case of CPM, effective dispersion dominates in a round-trip average approximation. This represents the dispersion integrated along the round-trip path and normalized by the cavity length. This effective dispersion, $GDD_{eff}$, is not a material property but an artificial parameter defined as the averaged dispersion per unit length under a round-trip average approximation. It can be expressed as:

$$GDD_{eff} = \frac{\partial^2 (\Delta k_F \cdot L + \Delta\varphi_C)}{\partial\omega^2}, \quad (2)$$

where $\Delta k_F$ is the wave vector mismatching in fiber, $\Delta\varphi_C$ is the phase compensation on the cavity mirror, L is the length of the fiber, $\omega$ is the corresponding angular frequency. This formula equates the phase matching of CPM to an effective dispersion parameter, which is fundamentally different from conventional dispersion definition. Moreover, the definition of $GDD_{eff}$ provides a unified framework for quantifying the phase compensation result within the cavity, thereby enabling direct comparison with conventional Kerr comb generation.

Here, we characterize the effective dispersion of the FFPR, in comparison with the natural dispersion of optical fiber. Although the overall phase mismatching inside the cavity, $\Delta\varphi_{FFPR} = \Delta k_F \cdot L + \Delta\varphi_C$ cannot be directly measured, it can determine indirectly via the relation GDD = $\partial^2\varphi/\partial\omega^2$. Therefore, the effective dispersion $GDD_{eff}$ can be experimentally characterized from the combined contributions of the dispersion in optical fiber and the dielectric coating. The dispersions of fiber and coating are respectively measured using interferometry (see details in Supplementary Information). As shown in Fig. 1c, the optical fiber with length of 10 mm (black curve) exhibits strong normal dispersion around 800 nm that fundamentally prevents soliton Kerr comb formation by disrupting the necessary phase matching. The dielectric coating exhibits anomalous dispersion (gray curve) in a wide wavelength range, tailoring the effective dispersion of FFPR. This effective dispersion is derived from the measured integrated dispersion, which is obtained by sweeping the frequency of a tunable laser and analyzing the resonant frequencies of the FFPR (Supplementary Fig. S4). The measurement (black points) reveals an effective anomalous dispersion region spanning over 10 nm. We quantify this using the second-order dispersion coefficient $D_2$[16,39], which we redefine here as $D_2 = (-c/n_0 D_1^2)\times(GDD_{eff}/L) = 5.54$ kHz. Also, the integrated dispersion profile demonstrates a tailored effective normal dispersion region below 790 nm. As a result, CPM creates both small effective normal and anomalous dispersion regimes from strong inherent material dispersion, thereby enabling phase matching for soliton Kerr comb.

The Q factor of FFPR can also be derived from its transmission when sweeping the frequency of a tunable laser. The resonance linewidth is measured to be 1.88 MHz at 790 nm (Fig. 1d), corresponding to a Q factor of $2.01\times10^8$. This high-Q factor attributes to CPM induced cavity enhancement effect, significantly reduces parametric oscillation thresholds, thereby enabling practical direct diode pumping.

Then we investigate Kerr comb generation and the experimental setup is shown in Fig. 2a. A frequency-doubled laser based on periodically-poled lithium niobate (PPLN) is built as pump source, with wide tunable wavelength range from 785 nm to 805 nm, covering both effective normal and anomalous dispersion regimes of the FFPR.

We scan the pump frequency across resonances in both the effective normal and anomalous dispersion regimes and measure the FFPR transmission with a pump power of 100 mW. For the effective normal dispersion case (Fig. 2b), we observe an approximately linear triangular resonance shape within 180 MHz laser detuning. This non-Lorentzian shape is caused by resonance thermal shifting. As for the effective anomalous dispersion case (Fig. 2c), distinct nonlinear dynamics emerge in the transmission signal. When laser detuning increases, the signal transitions from linear response to chaotic optical bursts—a signature of modulation instability (MI) in unstable Kerr comb states. When the laser enters red detuning from blue detuning, we observe a soliton step signature spanning 0.5 MHz detuning range. The identified soliton step corresponds to bright soliton formation from collapse of multiple unstable intracavity pulses in the effective anomalous dispersion regime, as shown in our simulations (Supplementary Fig. S5). Comparison between different effective dispersion regimes reveals significantly stronger MI manifestation under effective anomalous dispersion.

We then characterize Kerr comb spectrum evolution by slowly tuning the pump laser into the cavity resonance. When pumping at 788 nm, we observe Kerr comb generation in the small effective normal dispersion regime (Fig. 2d). The spectral evolution observed with increasing laser detuning features the emergence of new frequency components and progressive broadening centered at the pump wavelength. This spectral broadening leads to symmetrical peaks in the final Kerr comb spectra, a characteristic of small effective second-order normal dispersion. When pumping at 793 nm, we observe Kerr comb generation in the effective anomalous dispersion regime (Fig. 2e). The spectral evolution reveals that Kerr comb formation initiates with a primary set of comb lines characteristic of a Turing pattern. Subsequently, additional comb lines then emerge within the spectral gaps between these primary lines, eventually coalescing into a broader spectrum. The observed MI Kerr comb in the final spectra is enabled by effective second-order anomalous dispersion, which underpins broadband Kerr comb formation.

Tuning the pump wavelength across the effective anomalous dispersion region from 792 nm to 798 nm (Fig. 3a) yields further insights. At 798 nm, Kerr comb generation exhibits significant dispersive waves. Here, cascaded four-wave mixing generates thousands of comb lines, with dual dispersive waves emerging at 780 nm and 808 nm. The spectral positions of these waves align precisely with the two integrated zero-dispersion points measured in the cavity dispersion profile, confirming the tailored effective dispersion of FFPR by CPM. We detect a repetition rate signal exhibiting a single tone with significant noise, indicating that the Kerr comb remains in the MI state.

These results confirm an effective anomalous dispersion region from 792 nm to 798 nm is established through CPM.

Tuning the pump wavelength within this effective normal dispersion region from 788 nm to 785 nm (Fig. 3b) yielded a variety of distinct Kerr comb spectra. This spectral diversity arises due to the CPM design, which induces dramatic change in the effective dispersion. The measured repetition rate signal when pumping at 787 nm reveals a highly coherent Kerr comb state. Our simulations indicate that the effective dispersion at this wavelength supports the generation of dark solitons (Supplementary Fig. S5), exhibiting the characteristic dark-pulse temporal waveform. These results confirm small effective normal dispersion region around 787 nm is established through CPM.

Above findings show that CPM overcomes material dispersion limitations by redefining the effective dispersion of the cavity, and the low power requirement for low noise Kerr comb generation. Thus we further leverage the high-Q CPM device, to enable fully integrated system with direct diode pump, in a self-injection locking (SIL) setup. The SIL configuration substantially achieves low phase noise while streamlining soliton initiation dynamics. This approach ultimately facilitates robust, electrically driven Kerr comb operation in compact packages towards practical applications.

The internal structure and schematic diagram of the compact package are shown in Fig. 3c. This prototype synergizes a commercial laser diode, micro-optics components and an FFPR on the substrate, implemented in a 35×25 $mm^2$ footprint. Two micro-lens of different focal lengths are used to collimate and focus the laser diode output onto the FFPR, enabling entirely free-space optical coupling. Part of the circulating light in the cavity will re-emit out of the end-face establishing a self-injection locking feedback path to the laser diode. Active wavelength control is achieved through electrically controlled temperature stabilization.

We first characterize the self-injection laser performance within the compact package. Since the wavelength tuning range of a single diode laser cannot cover the full regions of effective normal and anomalous dispersion in FFPR, we developed two setups with same type of diode lasers but with different center wavelengths at 785 nm and 795 nm, respectively. The spectral characterization results of the self-injection lasers are shown in Fig. 3d. When operating without SIL, they exhibit prominent side modes spanning its gain bandwidth (blue trace). SIL (red trace) improves the side-mode suppression ratio to over 50 dB across a 20 nm spectral window. This spectral purification creates a clean pump background for Kerr comb generation. We also characterize the long-term wavelength stability of the SIL laser using the 785 nm setup as a representative example, as shown in Fig. 3e. The frequency drift decreases from 250 MHz/min (blue trace, without SIL) to ~1 MHz/min with SIL (red trace). Further high-resolution measurements of laser frequency stability and frequency noise are performed by mixing this laser with another narrow-linewidth 785 nm reference laser and analyzing the radio frequency (RF) beat note. The RF spectrum of the beat note signal (Fig. 4a) based on the same setup reveals a 5 MHz linewidth for the laser without SIL (blue trace) and a narrowed linewidth of 5 kHz with SIL (red trace). Additionally, we measured the frequency noise with and without SIL (Fig. 4b). The intrinsic linewidth with SIL is measured to be 17 Hz, indicating a linewidth reduction factor[40] of $3\times10^{5}$. Such strong noise suppression is mainly attributed to the high-Q factor of the FFPR. The Allan deviation (red points, inset of Fig. 3e) of laser frequency is measured to be $4\times10^{-9}$ at 1 s averaging time. This enhanced stability is crucial for maintaining long-term spectral coherence in Kerr comb generation. Together, these results demonstrate that self-injection locking significantly enhances laser diode performance. It simultaneously delivers narrow linewidth and exceptional wavelength stability, both of which are critical prerequisites for low-noise Kerr comb generation.

By strategically tuning the pump wavelength of the two Kerr comb packages to cover the effective normal and anomalous dispersion regimes of the FFPR, respectively, we generate various Kerr combs in both effective dispersion regimes (see details in Supplementary Information). When pumping at 787 nm, we observe a symmetric platicon Kerr comb (Fig. 4c), evidenced by its characteristic flat-top spectra with raised wings (red trace: experiment; blue trace: simulation). Time-domain simulations confirm rectangular pulse profiles, affirming switching wave formation. When pumping at 793 nm, we observe a bright soliton Kerr comb characterized by a $\mathrm{sech}^2$-like spectral envelope (Fig. 4d, red trace: experiment; blue trace: fit). Each of the Kerr comb exhibits high coherence, as evidenced by the narrow linewidth of their 10.1 GHz repetition rate signal measured with a high-speed GaAs photodetector (ET-4000F, EOT). The microwave signal is then sent to a phase noise analyzer (AnaPico, APPH20G) and a frequency counter (53230A, Keysight) to measure the single-sideband (SSB) phase noise and Allan deviation, respectively. The SSB phase noise power spectral density for platicon Kerr comb reaches -115 dBc/Hz at 10 kHz offset frequency and -125 dBc/Hz at 100 kHz offset frequency (Fig. 4e), whereas the bright soliton Kerr comb exhibits comparable phase noise at low offset frequency but higher noise floor at high offset frequency due to its lower optical power. Allan deviation measurement demonstrates the long-term stability of the repetition rate, which remains below $4\times10^{-8}$ at 1 s averaging time. This combination of low phase noise (<-125 dBc/Hz), high stability ($<4\times10^{-8}$) and compact integration satisfies requirements for field-deployable metrology applications.

Since our comb spectrum covers the Rb $D_1$ transition line, we can directly lock the comb to this atomic reference to further improve its long-term stability, which is necessary for metrology applications such as atomic physics and astronomical spectroscopy. Fig. 5a illustrates the experimental scheme. A tunable laser is locked on Rb $D_1$ transition line as reference (with frequency $f_{Rb}$ = 377.105 THz). We denote the central comb line of the packaged Kerr comb as $\mu_0$ (with frequency $f_{\mu0}$ = 378.336 THz), while the comb line nearest to the reference laser as $\mu_a$ (a = 122, $f_{\mu a}$ = 377.104 THz). The entire comb spectrum is frequency-shifted by an acousto-optic modulator (AOM), and the beat frequency between the comb line $\mu_a$ and the reference laser is locked to a rubidium-clock-referenced microwave source. The resulting error signal is fed back to the AOM RF driver, thereby stabilizing the entire Kerr comb (see details in Supplementary Information). We measure the frequency drifts of the central comb line $\mu_0$ using a wavemeter (WSU-2, resolution 2 MHz). In the free-running state, the frequency drift of comb line $\mu_0$ is more than 4 MHz over 30 minutes duration, as shown in Fig. 5b. After locking, the frequency drift of stabilized comb line $\mu_{0s}$ is greatly reduced as shown in Fig. 5c, with no observable drift beyond the wavemeter resolution range over the same time duration. The frequency drift of the reference laser is also plotted for better validating the locking performance. The performance of stabilized comb and reference laser are essentially identical, indicating that the stabilized comb has achieved a stability level comparable to that of the atomic reference. This result demonstrates the feasibility of directly referencing the CPM-based Kerr comb to an atomic transition, indicating that this new type of compact comb is promising to accelerate the practical applications of precision time-frequency metrology and astronomical spectroscopy.

In conclusion, we demonstrate CPM in a high-Q FFPR to achieve Kerr comb that can be directly stabilized to atomic reference. CPM overcomes the inherent material dispersion limitations, enabling phase matching in regimes where it is previously unattainable. Meanwhile, the cavity enhancement nature of CPM and the advanced device fabrication enable high-Q FFPR for low-phase-noise and low-threshold Kerr comb generation around 800 nm in a compact package. More importantly, the generated Kerr comb can be directly referenced to the rubidium $D_1$ transition so that the entire comb inherits the long-term stability of the atomic reference, showing significantly stability improvement compared to that in the free-running case. Together with the measured low phase noise of the 10.1 GHz repetition rate, this result establishes a compact, low-noise, and atomically referenced Kerr comb for precision metrology. Further optimization of CPM coating design will be essential to extend the phase-matched bandwidth. Such progress would allow broader spectral coverage of atomic and molecular transitions, further enhancing the versatility of CPM-based Kerr combs.

Beyond Kerr comb generation, CPM constitutes a transformative approach in nonlinear optics by enabling phase matching beyond the constraints of material and geometric dispersion, and is broadly applicable to diverse $\chi^{(3)}$ platforms[41]. By leveraging CPM, highly integrated nonlinear photonic devices with low operational thresholds can be realized across a wide spectral range, from the near-infrared to the visible and even ultraviolet. Such expansion in both platform diversity and operational bandwidth could drive significant advances in fields such as biomedicine, quantum optics, and nuclear optical clock. From a fundamental perspective, $\chi^{(3)}$ CPM establishes a synergy between artificial phase matching and cavity enhancement effect in third-order nonlinear systems, unlocking novel regimes for integrated nonlinear photonics.

## Methods

**Fiber Fabry-Perot resonator fabrication process and characterization**.

The FFPR in this work is made of few-mode fiber (from Yangtze Optical Fiber and Cable Joint Stock Limited Company). The main processing process of the FFPR is as follows: (1) Loading a full length of optical fiber into a ceramic ferrule then cut into 10 mm long individual resonator primitive embryo; (2) Hundreds of resonator primitives are mounted on a specially customized chuck and the two end faces are mechanically polished to ultra-high finish simultaneously; (3) Ultrasonic cleaning ensures that the two end surfaces where the optical film will be deposited are free of impurities; (4) Using $SiO_2$ and $Ta_2O_5$ as components, dispersive dielectric coating were deposited on two end faces of the polished resonator embryo. This coating has the characteristics of high reflectivity greater than 99.9% within 780 nm to 810 nm and negative group delay dispersion. The Q factor of the FFPR we fabricated is as high as 201 million at 790 nm, which is mainly attributed to the low propagation loss of the optical fiber and the wavelength-customizable optical thin film that maintains high reflectivity.

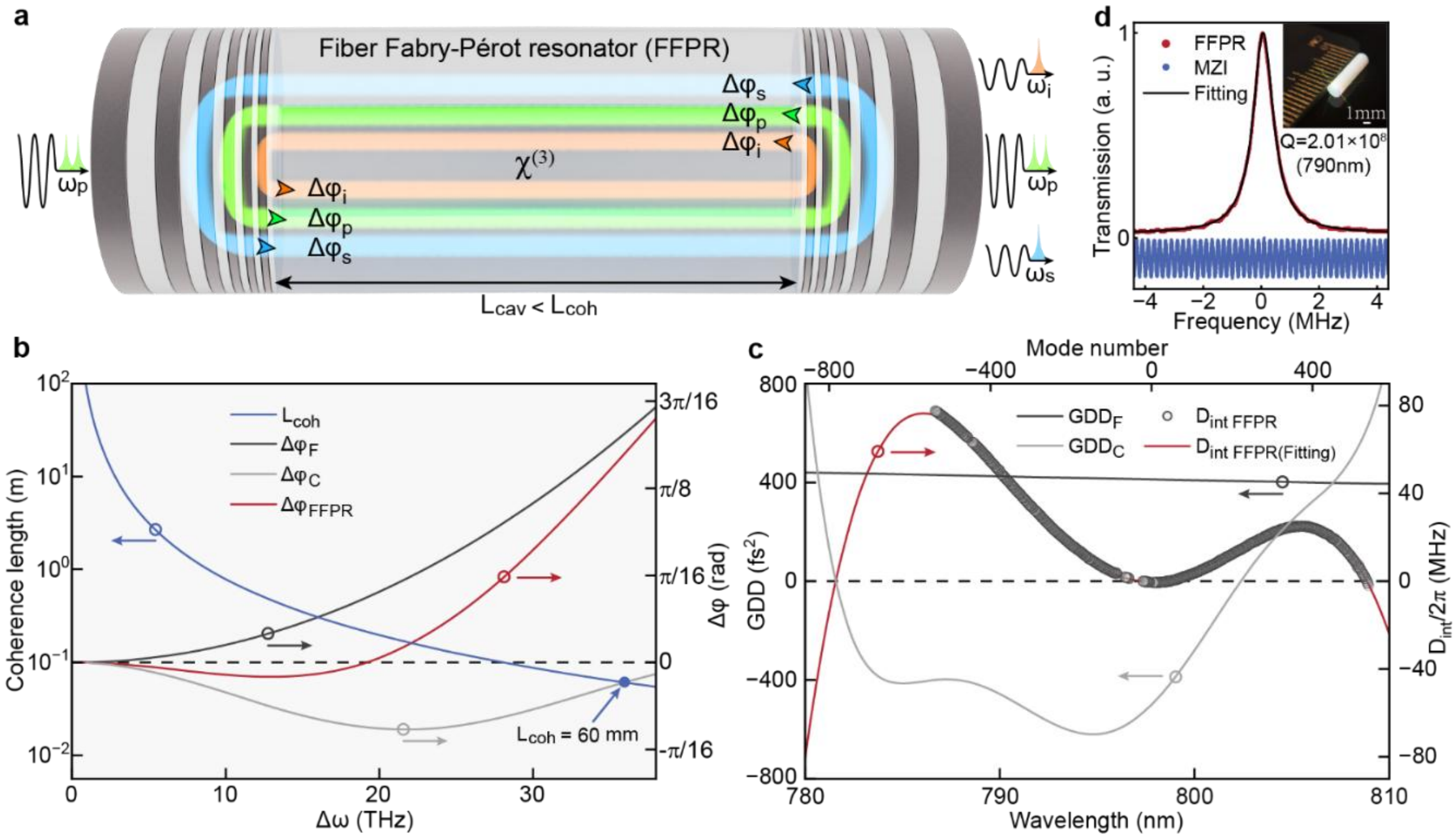

**Fig. 1|CPM for third-order nonlinearity within a macroscopic monolithic fiber resonator. a**, Schematic of $\chi^{(3)}$ CPM in an FFPR. Material dispersion induces phase mismatching for the resonant pump light $\omega_p$, signal light $\omega_s$, and idler light $\omega_i$. And the chirped multi thin-film coating on the cavity end face imparts compensating phase $\Delta\varphi_s$, $\Delta\varphi_p$, $\Delta\varphi_i$ that re-align their phases for achieving phase matching. By keeping the cavity length $L_{cav}$ smaller than coherence length $L_{coh}$, the nonlinear interaction length can be dramatically extended in the conjunction of resonant enhancement. **b**, Calculated coherence length in fiber is plotted in blue curve. For an FFPR with 10 mm length, the phase mismatching accumulated in fiber (black) is counteracted by dielectric coating (gray), resulting in a much smaller overall phase mismatching in the FFPR (red). **c**, Measured natural dispersion of fiber, artificial dispersion of coating and effective dispersion of FFPR. Through CPM, the intrinsic large normal dispersion (black) of fiber material is compensated by the artificial anomalous dispersion of coating (gray), resulting in an effective anomalous dispersion. This effective dispersion is confirmed by measured integrated dispersion (black points). The fitting curve (red) yields $D_2$ of 5.44 kHz. GDD, group delay dispersion. **d**, Transmission spectrum of the FFPR at 790 nm, showing a high Q factor of $2.01\times10^8$. A fiber Mach-Zehnder interferometer (MZI) is used to calibrate the frequency nonlinearity in laser scanning. Inset: photograph of 10-mm-long FFPR.

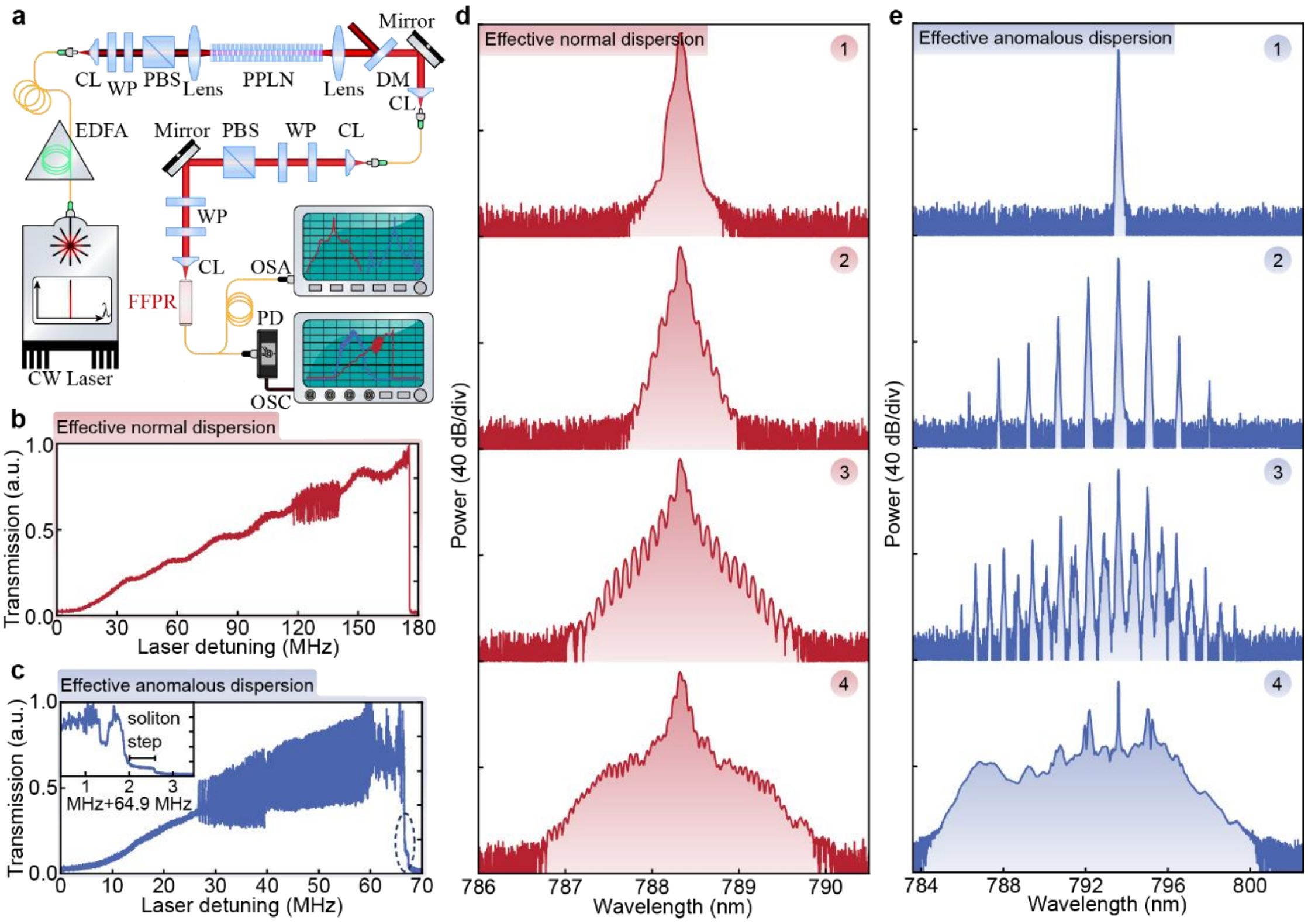

**Fig. 2|CPM-enabled Kerr comb evolution over both effective anomalous and normal dispersion regimes near 800 nm. a**, Experimental setup for near-infrared Kerr comb generation. Tunable CW pump laser is generated from PPLN frequency doubling. WP, wave plate; PBS, polarizing beam splitter; PPLN, periodically polarized lithium niobate; DM, dichroic mirror; CL, collimating lens; EDFA, erbium-doped fiber amplifier; FFPR, fiber Fabry-Perot resonator; PD, photodetector; OSA, optical spectrum analyzer; OSC, oscilloscope. **b**, Transmission spectrum of FFPR under laser frequency scanning in the effective normal dispersion regime. **c**, Transmission spectrum of FFPR under laser frequency scanning in the effective anomalous dispersion regime. Inset: the enlarged area shows a soliton step with laser detuning range of about 0.5 MHz. **d**, Kerr comb spectral evolution versus laser detuning in the effective normal dispersion regime. **e**, Kerr comb spectral evolution versus laser detuning in the effective anomalous dispersion regime.

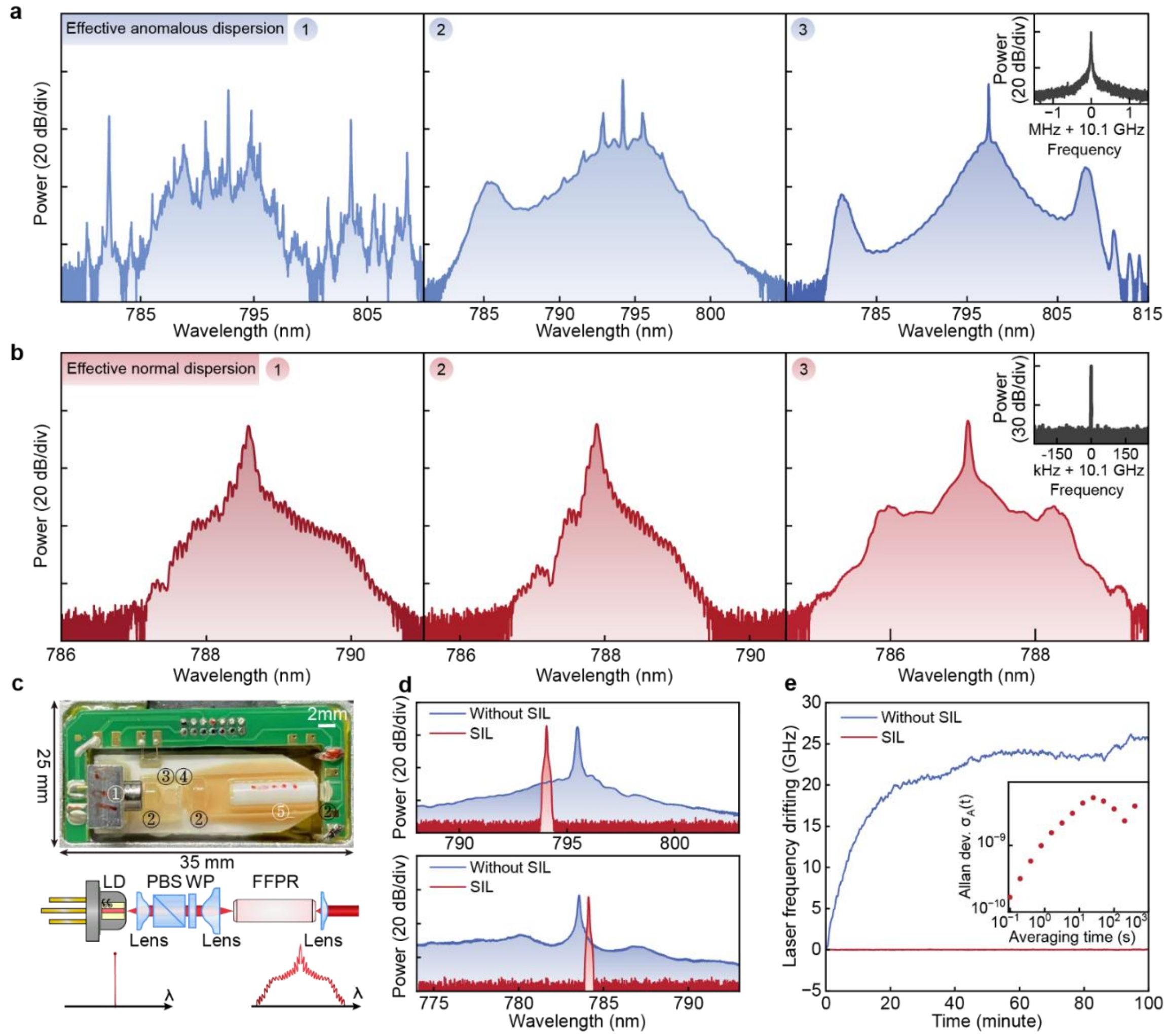

**Fig. 3|CPM-enabled Kerr comb spectra under different pump wavelengths and the packaged self-injection locking setup. a,** Kerr comb spectra under different pump wavelengths in the effective anomalous dispersion regime. Inset: RF beatnote centered at 10.1 GHz. RBW: 1 kHz. **b,** Kerr comb spectra under different pump wavelengths in the effective normal dispersion regime. Inset: RF beatnote centered at 10.1 GHz. RBW: 100 Hz. **c**, Schematic and photograph of the packaged self-injection locking setup. ①, LD, laser diode; ②, lens; ③, PBS, polarizing beam splitter; ④, WP, wave plate; ⑤, FFPR. **d**, Spectra of two different lasers with states of free running (blue) and self-injection locking (red). **e**, Laser frequency drifting with states of free running (blue) and self-injection locking (red). Inset: Allan deviation measurement of the self-injection locked laser frequency.

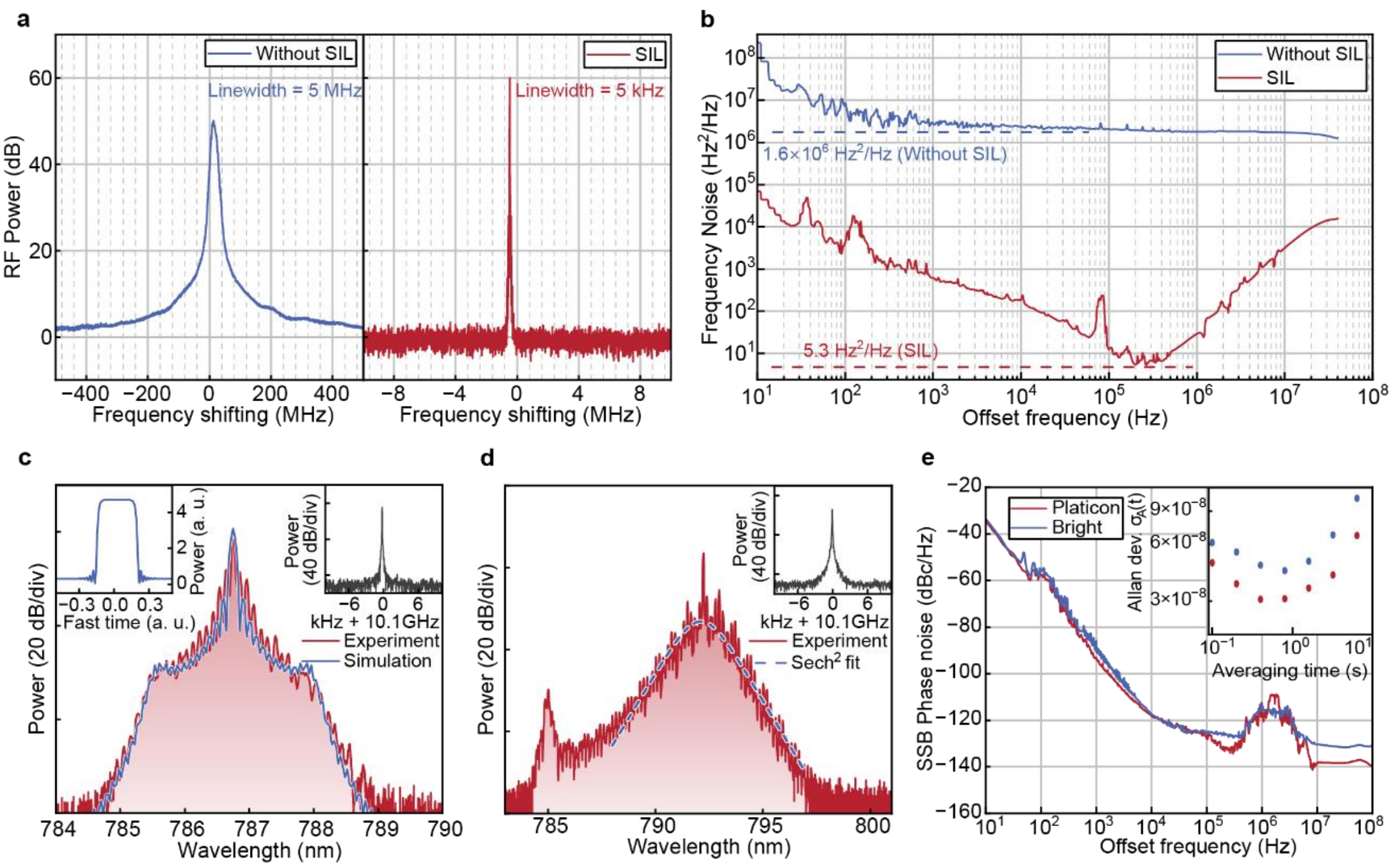

**Fig. 4|Low-noise Kerr combs generation around 785 and 795 nm in compact package. a**, Beating signals between the reference laser and the laser diode with states of free running (blue, RBW: 100 kHz) and self-injection locking (red, RBW: 2 kHz), showing linewidths of 5 MHz and 5 kHz, respectively. **b**, Laser frequency noise spectral densities with states of free running (blue) and self-injection locking (red). **c**, Experimental (red) and simulated (blue) results of platicon Kerr comb. Left inset: simulation of the time domain shows a rectangular optical pulse. Right inset: repetition rate signal of the platicon Kerr comb. RBW: 50 Hz. **d**, Experimental (red) and sech$^2$ fit (blue) of bright soliton Kerr comb. Right inset: repetition rate signal of the bright soliton Kerr comb. RBW: 50 Hz. **e**, Single sideband phase noise measurement of the platicon and bright soliton Kerr combs. Inset: Allan deviation measurement for 10.1 GHz repetition rate with a gate time of 0.1 s.

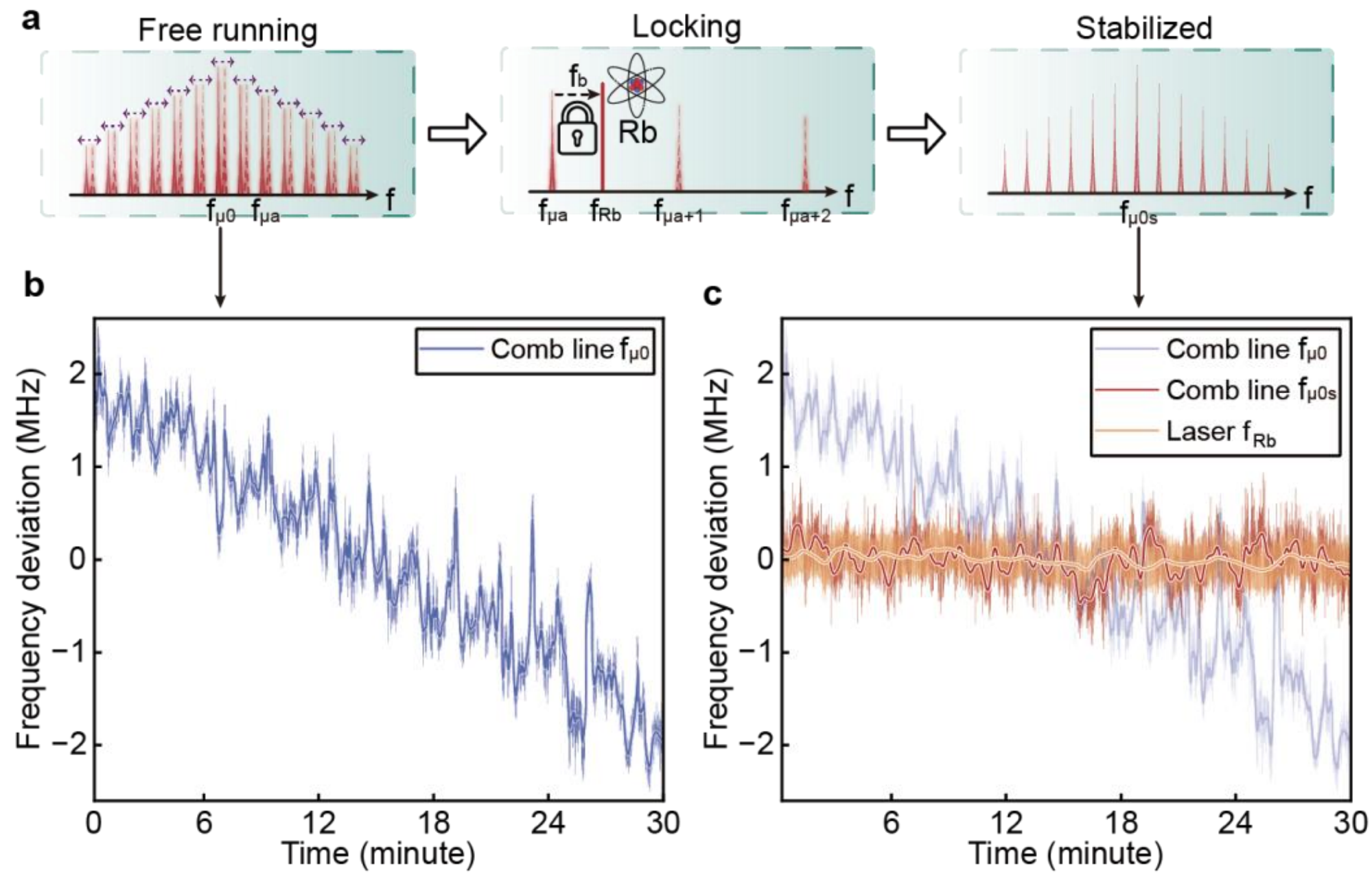

**Fig. 5|Frequency stabilization of a Kerr comb with reference to the rubidium $D_1$ transition line. a**, Experimental schematic for stabilizing the Kerr comb via Rb $D_1$ transition line. The free-running Kerr comb exhibits significant long-term frequency drift. By directly locking one comb line to a rubidium-stabilized laser, the frequency drift of whole comb can be greatly reduced. **b**, Measured frequency drift of the free-running Kerr comb. The blue line shows the frequency drift of comb line $\mu_0$ within 30 minutes. **c**, Measured frequency drift of the stabilized Kerr comb. The red and orange lines show the frequency drift of the stabilized comb line $\mu_{0s}$ and the rubidium-stabilized laser, respectively, within 30 minutes.

**Acknowledgements:** We appreciate N. Y. Zou for her contribution to the revision of this manuscript. This work was supported by the National Key R&D Program of China (2022YFA1205100), the National Natural Science Foundation of China (62288101, 62293523, 12304421, 12341403, 92463304, 92463308), Fundamental and Interdisciplinary Disciplines Breakthrough Plan of the Ministry of Education of China (JYB2025XDXM106), the Major Project of Scientific and Technological Innovation 2030 (2023ZD0301500), the Key project of Basic Research Program of Jiangsu Province (BK20253015), the Natural Science Foundation of Jiangsu Province (BK20230770, BK20232033), Fundamental Research Funds for the Central Universities (0213-14380292), CAS Project for Young Scientists in Basic Research (YSBR-69).

**Author contributions:** X. Y., K. J., W. L. and Z. X. formulated the original idea and designed the experiment. X. Y., K. J., J. J., W. L. prepared the CPM based FFPRs sample, designed and packaged the Kerr comb setup. X. Y., K. J. performed the measurements and simulations, as well as the data analysis. X. Y., K. J. and Z. X. wrote the manuscript. Z. Y., B. J. and Z. W. provided valuable feedback and advises. Z. X. and S. Z. supervised the whole work. All authors contributed to the manuscript preparation.

**Conflict of interests:** The authors declare no competing interests.

**Supplementary information** is available for this paper.

**Data and materials availability:** All data are available in the main text or the supplementary materials.

**Correspondence and requests for materials** should be addressed to Kunpeng Jia, Wei Liang and Zhenda Xie.